\documentclass[english,latin9,bibyear]{aa}
\usepackage[T1]{fontenc}
\usepackage{verbatim}
\usepackage{amsmath}
\usepackage{graphicx}

\makeatletter
\usepackage{natbib}
\usepackage[varg]{txfonts}
\usepackage{bm}

\usepackage{url}
\usepackage{natbib,twoopt}
\usepackage[breaklinks=true, colorlinks, citecolor=blue, linkcolor=magenta]{hyperref} 

\bibpunct{(}{)}{;}{a}{}{,} 
\makeatletter
\newcommandtwoopt{\citeads}[3][][]{\href{http://adsabs.harvard.edu/abs/#3}%
{\def\hyper@linkstart##1##2{}%
\let\hyper@linkend\@empty\citealp[#1][#2]{#3}}}
\newcommandtwoopt{\citepads}[3][][]{\href{http://adsabs.harvard.edu/abs/#3}%
{\def\hyper@linkstart##1##2{}%
\let\hyper@linkend\@empty\citep[#1][#2]{#3}}}
\newcommandtwoopt{\citetads}[3][][]{\href{http://adsabs.harvard.edu/abs/#3}%
{\def\hyper@linkstart##1##2{}%
\let\hyper@linkend\@empty\citet[#1][#2]{#3}}}
\newcommandtwoopt{\citeyearads}[3][][]%
{\href{http://adsabs.harvard.edu/abs/#3}
{\def\hyper@linkstart##1##2{}%
\let\hyper@linkend\@empty\citeyear[#1][#2]{#3}}}
\makeatother

\makeatother

\usepackage{babel}
\begin{document}

\title{}
\title{}
\title{The interstellar dust emission spectrum}
\subtitle{Going beyond the single-temperature grey body}
\author{François-Xavier Désert\inst{1}}
\institute{Univ. Grenoble Alpes, CNRS, IPAG, 38000 Grenoble France \\
\email{francois-xavier.desert@univ-grenoble-alpes.fr}}
\offprints{francois-xavier.desert@univ-grenoble-alpes.fr}
\date{Final version2, Received: 8 November 2021; Accepted: 13 December 2021}
\abstract{Most of the modelling of the interstellar dust infrared emission spectrum
is done by assuming some variations around a single-temperature grey-body
approximation. For example, the foreground modelling of Planck mission
maps involves a single dust temperature, $T$, along a given line-of-sight
with a single emissivity index, $\beta$. The two parameters are then
fitted and therefore variable from one line-of-sight to the other. }
         {Our aim is to go beyond that modelling in an economical way.}
         {We model the dust spectrum with a temperature distribution around
the mean value and show that only the second temperature moment matters.
We advocate the use of the
temperature logarithm as the proper variable.}
         {If the interstellar medium is not too heterogeneous, there is a universal analytical spectrum, which is derived here, that goes beyond the grey-body assumption. We show how the cosmic microwave background radiatively
interacts with the dust spectrum (a non-negligible corrective term
at millimetre wavelengths). Finally, we construct a universal ladder
of discrete temperatures, which gives a minimal and fast description
of dust emission spectra as measured by photometric mapping instruments
that lends itself to an almost linear fitting. This data modelling
can include contributions from the cosmic infrared background fluctuations.}{}

\maketitle

\section{Introduction}

When modelling the interstellar dust
emission it is usually assumed that its spectrum is close to that produced by a grey body. This assumption has been used over the last 50 years. Usually, the model
has been sufficient to explain the dust spectral energy distribution (SED)
along a given line-of-sight, be it in the diffuse interstellar medium
(ISM; \textit{e.g.} \citealp{Hensley2021}), in molecular clouds,
in star-forming regions, or in galaxies (\citealp{Galliano2018a}).
This SED is determined by a few broadband photometric measurements
with rather large uncertainties (statistical but also systematic),
so to the first order a single temperature is enough for the fits
to be satisfactory. The intensity, $I_{\nu}$, is thus modelled by
\begin{equation}
I_{\nu}=\tau_{0}\left(\frac{\nu}{\nu_{0}}\right)^{\beta}B_{\nu}(T_{d}),\label{eq:GreyBody}
\end{equation}where $\beta$ is the so-called emissivity index, $\tau_{0}$ is proportional
to the dust column density, and $B_{\nu}$ is the Planck function at
frequency $\nu$ and temperature $T$:
\begin{equation}
B_{\nu}(T)=\frac{2h\nu^{3}}{c^{2}}\frac{1}{\exp(\frac{h\nu}{kT})-1}\label{eq:BB1}
,\end{equation}where $h$ is the Planck constant, $c$ the light velocity, and $k$
is the Boltzmann constant.

In order to have a more physically motivated modelling of dust emission,
we can assume that the main dust component has different temperatures
along the line-of-sight. Studies have shown that the interstellar
radiation field (ISRF), as seen by a grain of dust, can be fluctuating due to shadowing or screening effects. That grain integrates
an absorption of the ISRF over $4\pi$ steradians; therefore, the average
temperature should be quite stable, though a fluctuation along that average
is unavoidable. 

For example, a single slab of attenuating dust placed in front of
the ISRF will produce grains at decreasing temperature $T_{d}=T_{0}e^{-\tau/(4+\beta)}$
as we enter the slab with opacity $\tau.$ The mass of grains at a
given temperature is thus $\frac{dm}{dT_{d}}=\frac{4+\beta}{T_{d}}\frac{dm}{d\tau}$,
an inverse power law of the temperature between the standard value,
$T_{0}$, and the value at the centre of the slab (hence a linear law
for the temperature logarithm).

Moreover, grains of different sizes will have slightly different temperatures.
While moving forward in complexity, we must not forget that the photometric
data for one pixel on the sky are scarce (e.g. nine values for observations
by the Planck mission in intensity). Therefore, we must find a flexible modelling
without too many parameters. It is also apparent that in many cases
a single-temperature and emissivity index model leads to false anti-correlations
between these parameters~\citep{Shetty2009}.

In this paper we investigate how to cope with the variability in
the dust temperature {along the line-of-sight} using Planck
function derivatives. Then we show how to take this effect through
a photometric mapping instrument. Considerations of the cosmic microwave
background (CMB) and the cosmic infrared background (CIB) are made
in this study.

\section{The model}

The simplest way to deal with a small temperature range is to assume
the dust temperature is distributed around a central value, $T_{m}$.
For reasons discussed below, we prefer to deal with the logarithm
variable $t\equiv\ln T$ (the normalisation is irrelevant in practice
as $t$ will only be used in a differential way). We can think of
the probability distribution as a normalised Gaussian function around
a mean value, $t_{m}=\ln T_{m}$, with a dispersion, $\sigma$, for a
given line-of-sight:
\begin{equation}
p(t)dt=\frac{1}{\sigma\sqrt{2\pi}}\exp\left(-\frac{(t-t_{m})^{2}}{2\sigma^{2}}\right)dt\label{eq:Gauss}
.\end{equation}
A single dimensionless parameter, $\sigma$, encapsulates the logarithm
temperature spread, and we assume $\sigma\ll1$. The Gaussian assumption
maximises the lack of information on the source of the spread, but
it will be shown that what matters is $\sigma$ as the second-order
moment of the distribution. The question becomes about the emission
spectrum of such a superposition of modified blackbodies. In the following
subsections, we develop the computation of such a spectrum using the
derivatives of the Planck function.

\subsection{Planck derivatives}

In order to simplify the computation, we define the dimensionless
reduced frequency as
\[
x\equiv\frac{h\nu}{kT}=\frac{h\nu}{ke^{t}}\,
\]

\noindent and the composite constant as $P\equiv\frac{2h}{c^{2}}$.

As such, one can reformulate Eq.~\ref{eq:BB1} as

\begin{equation}
B_{\nu}(T)=P\nu^{3}f(x)\,,\label{eq:BB2}
\end{equation}
with
\begin{equation}
f(x)=\frac{1}{e^{x}-1}\,,\label{eq:BB2-1}
\end{equation}
which is known as the photon occupation number. 

One can then compute the first and second derivatives of the Planck
function with the temperature logarithm as
\begin{equation}
\frac{dB_{\nu}}{dt}=P\nu^{3}g(x)\thinspace,\label{eq:dBdT}
\end{equation}
with
\begin{equation}
g(x)=-\frac{df}{d\ln x}=\frac{xe^{x}}{(e^{x}-1)^{2}},\label{eq:dBdT-1}
\end{equation}
and
\begin{equation}
\frac{d^{2}B_{\nu}}{dt^{2}}=P\nu^{3}g(x)(h(x)-1)\,,\label{eq:d2BdT2}
\end{equation}
with
\begin{equation}
\,\frac{d^{2}f}{d\ln x^{2}}=g(x)(h(x)-1)\,,\label{eq:d2BdT2-1}
\end{equation}
 and
\begin{equation}
h(x)=x\coth\frac{x}{2}=x\frac{e^{x}+1}{e^{x}-1}.\label{eq:d2BdT2-2}
\end{equation}

The Planck derivative functions are displayed in Appendix~1.

\subsection{Emission spectrum from a Gaussian temperature distribution}

We now proceed to combine the temperature spread, as written in Eq.~\ref{eq:Gauss},
with the Planck function to get the dust emission spectrum: 
\begin{equation}
I_{\nu}=\tau_{\nu}\int dt\,p(t)B_{\nu}(T),\label{eq:EmissionCombine1}
\end{equation}
where $\tau_{\nu}$ is the opacity along the line-of-sight. We now
make a Taylor expansion of the Planck function around a reference
logarithm temperature, $t_{0}=\ln T_{0}$:
\begin{align}
I_{\nu} & =\tau_{\nu}[B_{\nu}(T_{0})+\int dt\,p(t)(t-t_{0})\frac{dB_{\nu}}{dt}(T_{0})\nonumber \\
+ & \frac{1}{2}\int dt\,p(t)(t-t_{0})^{2}\frac{d^{2}B_{\nu}}{dt^{2}}(T_{0})].\label{eq:EmComb2}
\end{align}

We distinguish the reference temperature, $T_{0}$, from the Gaussian
distribution mean temperature as defined by $\ln T_{m}\equiv<\ln T>$
in order to be as general as possible. The integral over the temperature
can readily be separated from the frequency dependence such that we
obtain, by defining $\Delta t\equiv t_{m}-t_{0}$, 
\begin{align}
I_{\nu} & =\tau_{\nu}[B_{\nu}(T_{0})+\Delta t\,\frac{dB_{\nu}}{dt}(T_{0})\nonumber \\
+ & \frac{1}{2}\left(\Delta t^{2}+\sigma^{2}\right)\,\frac{d^{2}B_{\nu}}{dt^{2}}(T_{0})],\label{eq:EmComb3}
\end{align}
which, if $T_{0}=T_{m}$, simplifies into
\begin{equation}
I_{\nu}=\tau_{\nu}\left[B_{\nu}(T_{0})+\frac{\sigma^{2}}{2}\frac{d^{2}B_{\nu}}{dt^{2}}(T_{0})\right]\,.\label{eq:EmissionCombine4}
\end{equation}

This last expression shows how the dispersion in logarithmic temperature
translates into the addition of a new spectrum. This corresponds to
the second derivative of the Planck spectrum (Eq.~\ref{eq:d2BdT2})
times half the square of the dispersion (and times the emissivity). 

\begin{figure}
\includegraphics[width=0.45\paperwidth]{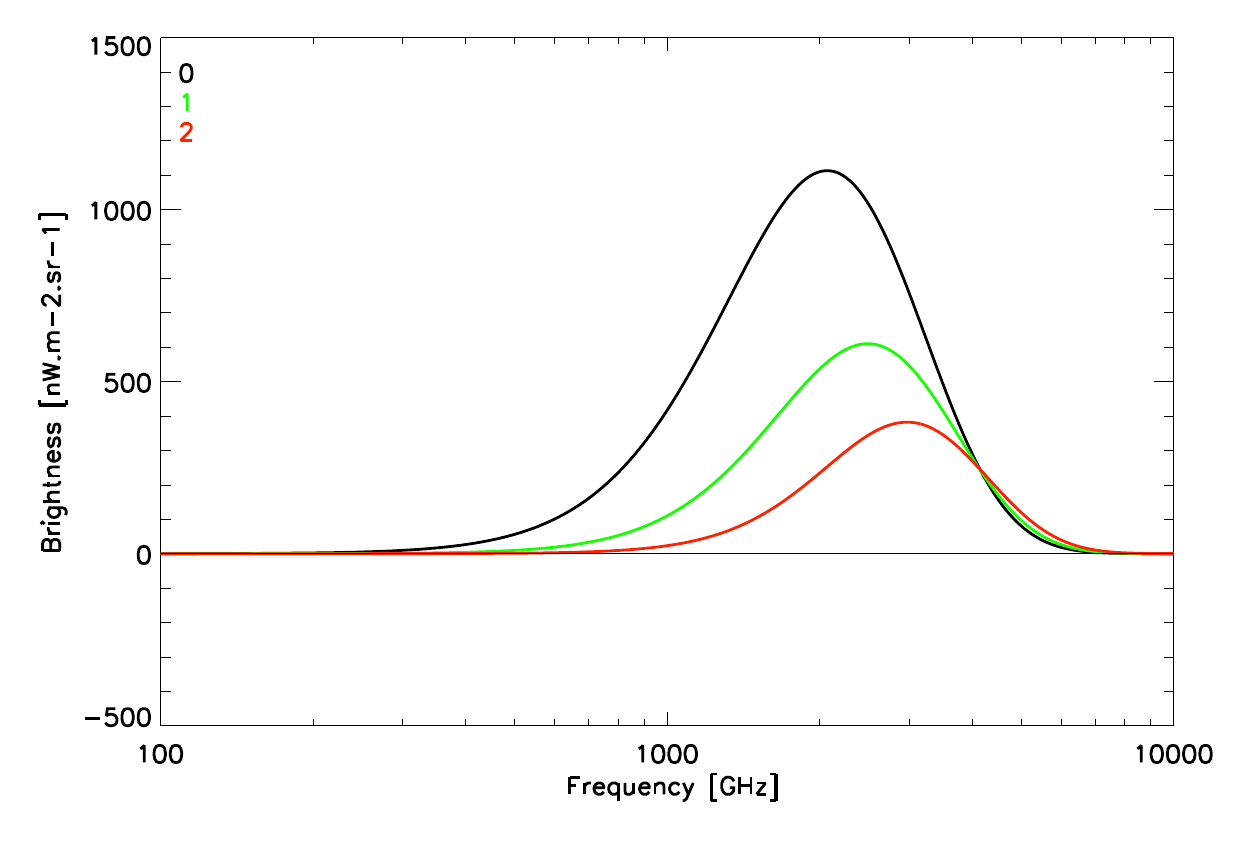}\caption{Three elementary spectra of dust {(see Eq.~\ref{eq:EmComb3})}
in a log-linear scale. For illustration purposes, we take the case
of 20~K dust with an emissivity of $10^{-4}(\frac{\nu}{353\,\text{GHz}})^{1.5}$.
The component of index~0 is the thermal brightness of dust. The component
of index~1 comes  from the first derivative of the Planck function (Eq.~\ref{eq:dBdT})
with $\frac{\Delta T}{T}=\Delta t=0.1$, and the component~2 is obtained
from the second derivative (Eq.~\ref{eq:d2BdT2}) using a logarithmic
temperature dispersion of $\sigma=0.15$.\label{fig:Spectrum}}
\end{figure}

The first-order derivative function disappears in the case of any
symmetrical temperature distributions like the Gaussian one. The three
spectra{l components of Eq.~\ref{eq:EmComb3}} are displayed
in the example of Fig.~\ref{fig:Spectrum}. While the component of
index~1 can be of both signs, the component of index~2 is always
positive for all frequencies and mostly influences the Wien
part of the spectrum. The three components have the same Rayleigh-Jeans
spectral behaviour in $\propto\nu^{2}T_{0}\tau_{\nu}$. The analogy
to the Sunyaev-Zel'dovich (SZ) effect, a CMB spectral distortion in clusters
of galaxies, is given in Appendix~2. The complete formalism of the interaction
of a temperature distribution with the Planck function was developed
by \citet{Stebbins2007}. He shows that logarithm temperature moments
are the key quantities to know, irrespective of the exact distribution, here $p(t)$, as can be seen in our Eq.~\ref{eq:EmComb2}. \citet{PitrouStebbins2014}
have argued for $t\equiv\ln T$ as the proper variable to use in the
context of the CMB distortions (including relativity Lorentz frame
invariance). We think that there is an additional reason for that
use that is linked to the way photometric data are spread in frequency
(see below).

\subsection{Finding the underlying distribution of dust temperature from photometry}

We now see the interest in the decomposition of the dust emission
spectrum into the Planck function and its two derivatives (Eq.\,\ref{eq:EmComb3}).
With only a few parameters, we can describe quite different situations.
Photometric measurements are used to deduce dust properties,
abundance, and luminosity. They are obtained from broadband detectors: 
photoconductors, bolometers, and kinetic inductance detectors (KIDs). The width of the band $n$, of
a given instrument and telescope, described by the transmission curve
$H_{n}(\nu)$, is so large that one has to assume a reference spectrum,
$J_{\nu}$, to quote a brightness, $I_{\nu}^{n}$, at the band central
frequency, $\nu_{n}$, and for a given underlying spectrum, $I_{\nu}$:
\begin{equation}
I_{n}=\frac{J_{\nu_{n}}}{\int d\nu\thinspace H_{n}(\nu)\thinspace J_{\nu}}\int d\nu\thinspace H_{n}(\nu)\thinspace I_{\nu}\thinspace.\label{eq:Bandpass}
\end{equation}

To simplify further, we define the set of normalised transmission
functions for a given instrument as
\[
G_{\nu}^{n}=\frac{J_{\nu_{n}}}{\int d\nu\thinspace H_{n}(\nu)\thinspace J_{\nu}}H_{n}(\nu)\thinspace,
\]
and we can rewrite the bandpass Eq.~\ref{eq:Bandpass} as:
\begin{equation}
I_{n}=\left\langle G_{\nu}^{n}.I_{\nu}\right\rangle,\label{eq:Bandpass2}
\end{equation}
where the bracket indicates a simple integration over the frequencies.

The reference spectrum, $J_{\nu}$, is usually taken as the first derivative
of the CMB Planck function for millimetre wavelengths (including
the 353~GHz Planck mission band),
\[
J_{\nu}=\frac{P\nu^{3}}{T_{CMB}}g(\frac{h\nu}{kT_{CMB}}),
\]
 and the $J_{\nu}\propto\nu^{-1}$  Infrared Astronomical Satellite (IRAS) convention at submillimetre
wavelengths. The ratio $I_{\nu}(\nu_{n})/I_{n}$ is close to one and
is usually called the colour correction. 

The bandpass $H_{n}$ has a typical width of $\Delta\nu_{n}$ and
is usually characterised by a resolution of $\frac{\nu_{n}}{\Delta\nu_{n}}\sim3$
irrespective of $\nu_{n}$. So the effective spectral resolution
is coarse but constant in logarithm terms. It is thus appropriate
to match this logarithmic scale with the temperature grid that we
seek. For that purpose, it is worth remembering that dust cannot be
colder than the CMB. Therefore, we propose discretising the temperature
as follows:
\begin{equation}
t_{l}=\log T_{l}=\log T_{CMB}+l.s,\label{eq:GridTemp}
\end{equation}
or $T_{l}=T_{CMB}.e^{l.s}$, where $l$ is an integer starting at
0 for the CMB temperature. The logarithmic temperature step, $s$, should
match the photometric gridding, so values of $\frac{1}{3}$ or $\frac{1}{2}$
seem appropriate. The grid upper limit depends on the chosen experiments.
Defining $x_{l}\equiv\frac{h\nu}{kT_{l}}$, and using Eq.~\ref{eq:EmComb3},
we can now rewrite the dust brightness as a linear decomposition of
known functions that are fixed by the grid of Eq.~\ref{eq:GridTemp}:
\begin{align}
I_{\nu} & =P\nu^{3}\sum_{l=0}\tau_{\nu}^{l}[f(x_{l})+\Delta t_{l}\,g(x_{l})+\,\nonumber \\
 & \frac{1}{2}\left(\Delta t_{l}^{2}+\sigma_{l}^{2}\right)\,g(x_{l})\,(h(x_{l})-1)]\,,\label{eq:GridInu}
\end{align}
where the $l$ component is described by its opacity law, $\tau_{\nu}^{l}$,
and its temperature distribution, which is solely characterised by
two parameters: the logarithm mean temperature shift, $\Delta t_{l}=t_{l}^{m}-t_{l}$,
and the dispersion, $\sigma_{l}$. In order to compare the model with
observations, we must take into account the passage
of light through an instrument, by combining Eq.~\ref{eq:GridInu}
and Eq.~\ref{eq:Bandpass2}. For that purpose, we normalise the opacity
$\tau_{\nu}^{l}$ to a reference opacity, $\tau_{l}$ (say, at a given
frequency). We can then define three series of coefficients, which
describe the coupling between the dust temperature grid and the set
of photometric bands:
\begin{align*}
K'_{0,n,l} & \equiv\left\langle G_{\nu}^{n}.\left[P\nu{{}^3}\frac{\tau_{\nu}^{l}}{\tau_{l}}f(x_{l})\right]\right\rangle \,,
\end{align*}
\begin{align*}
K_{1,n,l} & \equiv\left\langle G_{\nu}^{n}.\left[P\nu{{}^3}\frac{\tau_{\nu}^{l}}{\tau_{l}}g(x_{l})\right]\right\rangle \,,
\end{align*}
\begin{align}
K_{2,n,l} & \equiv\left\langle G_{\nu}^{n}.\left[P\nu{{}^3}\frac{\tau_{\nu}^{l}}{\tau_{l}}g(x_{l})(h(x_{l})-1)\right]\right\rangle \,,\label{eq:Coeff}
\end{align}
and we obtain the photometric expectation in band $n$:
\begin{align}
I_{n} & =\sum_{l=0}\tau_{l}[K'_{0,n,l}+\Delta t_{l}K_{1,n,l}+\frac{1}{2}\left(\Delta t_{l}^{2}+\sigma_{l}^{2}\right)K_{2,n,l}]\,.\label{eq:Global1}
\end{align}

We now need a digression with respect to the first temperature bin
at $T_{CMB}$, which is going to modify $K'$.

\subsection{The radiative transfer of the CMB through dust}

We now want to take into account the {dominant} background,
which goes through interstellar dust before reaching the instruments\footnote{ This study was reported to the Planck
mission consortium in 2014 but was never published.}. It happens that dust, although a very thin layer with opacities
much smaller than 1, has a non-negligible impact on the CMB and vice versa.
The radiative transfer equation is solved for a single-temperature
dust to get
\[
I_{\nu}=B_{\nu}(T_{CMB})+(1-e^{-\tau_{\nu}})\left[B_{\nu}(T_{d})-B_{\nu}(T_{CMB})\right],
\]
which can be rewritten, for the optically thin case, as the real modified
blackbody (RMBB) law:
\begin{equation}
I_{\nu d}=\tau_{\nu}\left[B_{\nu}(T_{d})-B_{\nu}(T_{CMB})\right],\label{eq:RealMBB}
\end{equation}
where we have suppressed the CMB monopole term, which is cancelled out in
the measurement by the current differential experiments, except the Far Infrared Absolute Spectrophotometer (FIRAS) on the Cosmic Background Explorer (COBE). The RMBB law is important in the millimetre domain. For
example, at $100\thinspace\text{GHz}$ the correction amounts to
a 7\% downward effect for $18\,\text{K }$ dust, irrespective of the
emissivity law. The RMBB can be compared with the occultation of the
CMB by planets, which is taken into account in the calibration of
cold planets. Equation~\ref{eq:RealMBB} is equivalent to saying that
in Eq.~\ref{eq:Global1} the first term, $K'_{0,n,l}$, is special
and has to be modified into $K_{0,n,l}\equiv K'_{0,n,l}-K'_{0,n,0}$
so that very cold dust (at the CMB temperature) becomes invisible
(a consequence of Kirchhoff's law). The influence of the CMB
on the submillimetre spectrum of high-redshift galaxies has also been
noted by \citet{daCunha2013}. Here we show that, independent of redshift,
the CMB has to be explicitly included in the dust SED.

We can now rewrite Eq.~\ref{eq:Global1} as 
\begin{align}
I_{n} & =\sum_{l=0}\tau_{l}[K_{0,n,l}+\Delta t_{l}K_{1,n,l}+\frac{1}{2}\left(\Delta t_{l}^{2}+\sigma_{l}^{2}\right)K_{2,n,l}]\,.\label{eq:Global2}
\end{align}
Although $K_{0,n,0}=0$, very cold dust can still manifest itself with the other two CMB terms:
$K_{1,n,0}$ and $K_{2,n,0}$. We note that it is more likely coming
from the CIB than from interstellar dust. \citet{Elfgren2004} have made
a case for part of the CIB being in the spectral form encapsulated
by the coefficients $K_{1,n,0}$.

\section{Discussion}

We think that the result described in Eq.~\ref{eq:Global2} is
general enough to encompass the description of many dust environments.
It includes the modified blackbody (MBB) case if there is only one $l$ component, without dispersion
($\sigma_{l}=0)$. The data at hand in photometric bands
$I_{n}^{meas}$ (with the associated error bar $\sigma_{n}^{meas}$) can be easily compared with this model. If we assume a dust emissivity
law (identical or not for different $l$ components), the coefficients
$K$ can be pre-computed once and for all, thus making the fitting
process very fast for many lines of sight: 
\begin{equation}
\chi^{2}=\sum_{n}\left(I_{n}^{meas}-\sum_{l=0}\sum_{i=0}^{2}c_{i,l}K_{i,n,l}\right)^{2}/(\sigma_{n}^{meas})^{2}\,.\label{eq:chi2}
\end{equation}
If we keep only the first two coefficients, $K_{0,n,l}$ and $K_{1,n,l}$,
the fit is strictly linear, with the condition that $c_{0,l}$ must
be positive. The number of components, as described by the number
of useful $l$ components, must be tuned in the process in order to
keep $\Delta t_{l}\equiv t_{l}^{m}-t_{l}=\frac{c_{1,l}}{c_{0,l}}$
small. In that sense, the process is parsimonious: finding one $l$ component
that is best at minimising the $\chi^{2}$ and testing to see if a second one
is really needed, then iterating. If we add the temperature distribution
with $K_{2,n,l}$ and $\sigma_{l}$ , and compare Eq.~\ref{eq:Global2}
with Eq.~\ref{eq:chi2}, we can recover a linear fit if we include
Lagrange coefficients to satisfy the condition $c_{2,l}\ge\frac{c_{1,l}^{2}}{2c_{0,l}}$
.

The true non-linear part of the fit is in the emissivity law, which
is hidden in the $K$ coefficients. What is also hidden is the absolute
calibration uncertainty of the instrument. This second set of values
is degenerate with $K$. Another complication arises with the emissivity
law, which is, in principle, different for different $l$. One can
assume the same emissivity for a starter and refine that hypothesis
if it happens to be really necessary. From Fig.~\ref{fig:Spectrum}
we see that a temperature spread always produces a positive distortion
at all frequencies. The effect is that an MBB fit will tend to overestimate
the dust temperature and to undervalue the emissivity index. So
not only do we have statistically anti-correlated error bars between
these two quantities (\citealp{Shetty2009}), but the (genuine) temperature
spread along the line-of-sight can cause an apparent $T$-$\beta$
anti-correlation too. 

One could find that this model has `too many'{}
parameters. But we would argue that for the case $s=\frac{1}{3}$,
this corresponds to only five discrete temperatures between 8 and 30~K,
namely 9.1, 12.2, 16.5, 22.3, and 30~K. Secondly, the fitting procedure
can cull the components as the parameters in Eq.~\ref{eq:chi2} have
to conform to positivity constraints set by Eq.~\ref{eq:Global2}.
With the case shown in Fig.~\ref{fig:Spectrum}, a single-temperature
logarithmic change by $\frac{s}{2}$ yields a maximum residual of
$0.6\,\%$ relative to the peak emission for three components fitted
with Eq.~\ref{eq:chi2} ($3.5\,\%$ if only two components are used).

Another constraint will show up when one finds that the temperature
spread is larger than the bin size, $s$. In that case, the parameter
$\sigma$ should saturate to $\frac{s}{2\sqrt{3}}$ (the equivalent
of uniform noise in a digital binning process), and hence $\sigma_{sat}^{2}\sim0.009\left(\frac{s}{0.33}\right)^{2}$.

The bolometric luminosity of dust can be directly computed from a
linear combination of integrals pre-computed from Eq.~\ref{eq:GridInu}.
Polarisation photometry follows the same rules as intensity. We can
expect the degree of dust polarisation to be strong when the
temperature distribution is not too spread out (\textit{i.e. }for
small $\sigma$).

The CIB fluctuations \citep{Planck_2013_30_CIB} may be revisited
in light of this model {because of the linearisation procedure},
as it can lend itself to the powerful statistical tools used for CMB
studies, including covariances and model testing with likelihood functions,
via the statistics of the $c_{i,l}$ coefficients. 

{This work is complementary to the approach by \citet{Chluba2017}.
Both studies put the emphasis on Planck derivatives and temperature
moments. \citet{Chluba2017} argues for $\frac{1}{T}$ as the proper
expansion variable. We prefer $\ln T$ because it will fare better,
in terms of the number of required steps in the temperature grid,
for broad temperature distributions (including the CIB and the example
in Sect.~1), where more than one $l$ component is necessary. In
that case, high-frequency convergence problems, signalled by \citet{Chluba2017},
are also alleviated. Moreover, for broadband experiments, it is impossible
to go beyond the second Planck derivative as the fitting system becomes
too under-determined for limited signal-to-noise ratio measurements.}

\section{Conclusions}

We have shown an analytical development around the Planck function.
We have emphasised the role of its second derivative with respect to
the logarithm of the dust temperature. This is important in order to
include the effect of a very likely distribution of temperatures along
the line-of-sight around the main dust temperature. We have also
devised an economical way of accounting for a wider temperature range
by discretising the temperature ladder in constant logarithm steps,
which naturally starts at the CMB temperature. By explicitly including
the instrumental measurement process, we have devised a model that
can be fit in an almost linear way, from photometry to dust
temperature distributions, thus accelerating the computation of the
inverse problem. Further work will have to be done to implement these
findings for the analysis of dust in the Planck and Herschel missions,
in particular the optimal gridding of dust temperatures versus the
experimental photometric frequency sampling.  This work is
complementary to investigations of the possible variations in the
emissivity index in some galactic regions
\citep{Mangilli2021,Rigby2018,Bracco2017,Tang_CMZ_2021,PlanckCollab_PIP14_2014}.
Applications to the CIB statistics and dust as the major foreground
for CMB studies should be sought too.
\begin{acknowledgements}
The author thanks discussions with François-Xavier Hamel, Guilaine
Lagache and Nicolas Ponthieu.
\end{acknowledgements}

\bibliographystyle{aa}
\bibliography{biblio}

\begin{appendix}

\section{Planck derivative functions}

The third Planck derivative is useful for assessing the errors made by
using only two derivatives in the Taylor expansion. For completeness,
it is given here: 
\begin{equation}
\frac{d^{3}B_{\nu}}{dt^{3}}=P\nu^{3}g(x)\left[(h(x)-1)^{2}-h(x)+2x\,g(x)\right]\,.\label{eq:3rdDerivative}
\end{equation}

\begin{figure}
\includegraphics[width=0.45\paperwidth]{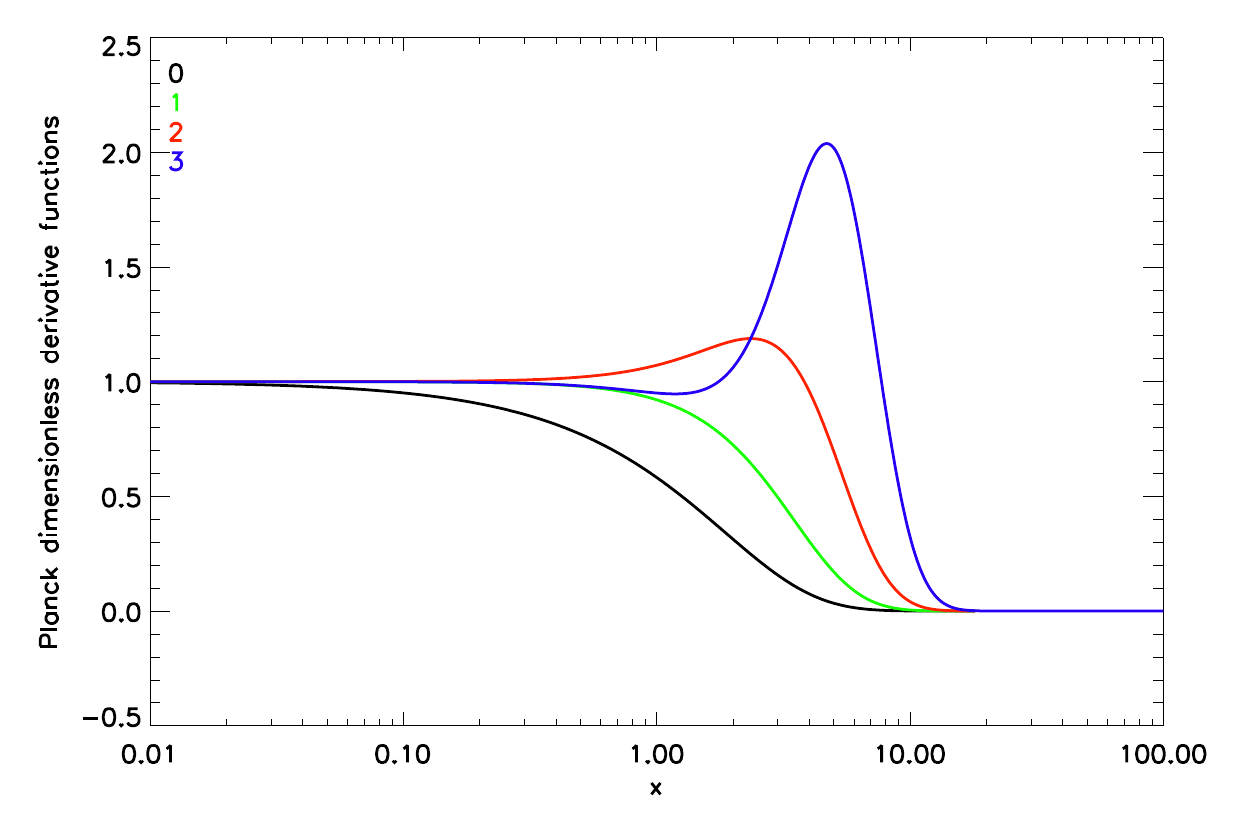}\caption{Four Planck functions (times $x$) as a function of the dimensionless
frequency parameter $x$. The Planck function has the 0 index, and the three indices label the three consecutive Planck function derivatives.
Hence, we have plotted $x\,f(x)$, $x\,g(x)$, $x\,g(x)\,(h(x)-1)$,
and $x\,g(x)\,\left[(h(x)-1)^{2}-h(x)+2x\,g(x)\right]$, where the
functions $f$, $g$ , and $h$ are defined in Eqs.~\ref{eq:BB2-1},
\ref{eq:dBdT-1}, and \ref{eq:d2BdT2-2}. \label{fig:PlDerFunctions}}
\end{figure}
It is worth showing the graph of Planck derivative functions. As they
all diverge at low frequencies as $\frac{1}{x}$~, we can show instead
$x$ times the functions. In Fig.~\ref{fig:PlDerFunctions} we show
the Planck function and its three derivatives (with the index going from
0 to 3).

\section{The Sunyaev-Zel'dovich effect}

The second derivative of the Planck function is linked to the SZ effect, a spectral distortion of the CMB through clusters of
galaxies. The SZ effect is quantified by the Compton parameter, $y$,
which is proportional to the integrated thermal electron pressure
along the line-of-sight. Indeed, if we specifically perform the substitution
$\Delta t=-3y$ and $\sigma^{2}=2y$ in Eq.~\ref{eq:EmComb3},$ $
where $T_{0}=T_{CMB}$ is the present CMB
temperature \citep{Stebbins2007}, we recover the SZ non-relativistic
thermal distortion:
\begin{equation}
I_{\nu}^{tSZ}=y\,P\nu^{3}g(x)(h(x)-4)\,.\label{eq:SZ}
\end{equation}
 This gives a direct interpretation of the SZ distortion: the Compton
effect (while maintaining the number of photons constant) shifts the
CMB average temperature along the cluster line-of-sight by a relative
factor of $-3y$, and the Maxwellian electron velocity distribution
interacts with the CMB photons as if they produced a Gaussian temperature
distribution with a dispersion $\sigma.$ This explains why the quadratic
equivalent temperature dispersion in logarithm is proportional to
the quadratic Maxwellian velocity distribution (\textit{i.e. }the
electron temperature, $T_{e}$): $\sigma^{2}=2\frac{kT_{e}}{m_{e}c^{2}}\tau$,
where $m_{e}$ is the electron mass and $\tau$ is the line-of-sight
opacity to the Compton effect.

One of the differences between the SZ spectral distortion (in $h-4$) and
the Planck second derivative (in $h-1$) is that the SZ effect has
a negative relative distortion in the Rayleigh-Jeans part of the spectrum,
the so-called SZ decrement: $\frac{\Delta I_{\nu}}{I_{\nu}}\simeq-2y$. 

\end{appendix}
\end{document}